\documentclass[10pt, conference]{IEEEtran}
\IEEEoverridecommandlockouts
\usepackage{cite}
\usepackage{amsmath,amssymb,amsfonts}
\usepackage{algorithmic}
\usepackage{graphicx}   
\usepackage{textcomp}
\usepackage{xcolor}
\usepackage{tikzscale}
\usepackage{booktabs,url}
\usepackage{tabularx}
\usepackage{balance}
\usepackage{multirow}
\usepackage{graphics}
\usepackage[list=true]{subcaption}
\usepackage[font={footnotesize}]{caption}
\renewcommand{\baselinestretch}{1.01}

\def\BibTeX{{\rm B\kern-.05em{\sc i\kern-.025em b}\kern-.08em
    T\kern-.1667em\lower.7ex\hbox{E}\kern-.125emX}}
\usepackage{acro}
\usepackage{pgfplots}
  \pgfplotsset{compat=newest}
  \usetikzlibrary{plotmarks}
  \usetikzlibrary{arrows.meta}
  \usepgfplotslibrary{patchplots}
  \usepackage{grffile}
  \usepackage{amsmath}
  \pgfplotsset{plot coordinates/math parser=false}
  \newlength\figureheight
  \newlength\figurewidth
\DeclareAcronym{5g}{
short=5G,
long= fifth generation,
}
\DeclareAcronym{dsp}{
short=DSP,
long= digital signal processing,
}
\DeclareAcronym{nn}{
short=NN,
long= neural network,
}
\DeclareAcronym{mlp}{
short=MLP,
long=multilayer perceptron
}
\DeclareAcronym{GaN}{
short=GaN,
long=Gallium Nitride,
}
\DeclareAcronym{relu}{
short=ReLU,
long = rectified linear unit, 
}
\DeclareAcronym{mse}{
short=MSE,
long=mean squared error,
}
\DeclareAcronym{rvtdnn}{
short=RVTDNN,
long= real-valued time-delay neural network,
}
\DeclareAcronym{arvtdnn}{
short=ARVTDNN,
long= augmented real-valued time-delay neural network,
}
\DeclareAcronym{r2tdnn}{
short=R2TDNN,
long= residual real-valued time-delay neural network,
}
\DeclareAcronym{flops}{
short=FLOPs,
long= floating point operations,
}
\DeclareAcronym{ofdm}{
short=OFDM,
long=orthogonal frequency division multiplexing,
}
\DeclareAcronym{par}{
short=PAR,
long=peak-to-average ratio,
}
\DeclareAcronym{papr}{
short=PAPR,
long=peak-to-average power ratio,
}
\DeclareAcronym{rf}{
short=RF,
long=radio frequency,
}
\DeclareAcronym{pa}{
short=PA,
long=power amplifier,
}
\DeclareAcronym{pas}{
short=\acs{pa}s,
long=power amplifiers,
}
\DeclareAcronym{psd}{
short=PSD,
long= power spectral density,
}
\DeclareAcronym{dpd}{
short=DPD,
long=digital predistortion,
}
\DeclareAcronym{cfr}{
short=CFR,
long=crest factor reduction,
}
\DeclareAcronym{cf}{
short=CF,
long=crest-factor}

\DeclareAcronym{evm}{
short=EVM,
long=error vector magnitude,
}
\DeclareAcronym{nmse}{
short=NMSE,
long=normalized mean square error,
}
\DeclareAcronym{acpr}{
short=ACPR,
long=adjacent channel power ratio,
}
\DeclareAcronym{pae}{
short=PAE,
long=power added efficiency,
}
\DeclareAcronym{dla}{
short=DLA,
long=direct learning architecture,
}
\DeclareAcronym{ila}{
short=ILA,
long=indirect learning architecture,
}
\DeclareAcronym{ilc}{
short=ILC,
long=iterative learning control ,
}
\DeclareAcronym{cfr-dpd}{
short=CFR-DPD,
long=CFR combined with DPD,
}
\DeclareAcronym{icf}{
short=ICF,
long=iterative clipping and filtering,
}
\DeclareAcronym{am/am}{
short=AM/AM,
long=amplitude-to-amplitude,
}
\DeclareAcronym{am/pm}{
short=AM/PM,
long=amplitude-to-phase,
}
\DeclareAcronym{mimo}{
short=MIMO,
long=multiple-input multiple-output
}
\DeclareAcronym{mp}{
short=MP,
long=memory polynomial
}
\DeclareAcronym{gmp}{
short=GMP,
long=generalized memory polynomial
}
\DeclareAcronym{adc}{
short=ADC,
long= analog-to-digital converter}
\DeclareAcronym{dac}{
short=DAC,
long= digital-to-analog converter}
\DeclareAcronym{ilc-dpd}{
short=ILC-DPD,
long= adaptive ILC-based DPD
}
\DeclareAcronym{rms}{
short=RMS,
long= root mean squares
}
\DeclareAcronym{vst}{
short=VST,
long= vector signal transceiver
}
\DeclareAcronym{mmwv}{
short=mm-Wave,
long= millimeter-wave
}

\begin{document}

\title{Residual Neural Networks \\for Digital Predistortion \\
\thanks{This work was supported by the Swedish Foundation for Strategic Research (SSF), grant no.~I19-0021.}
}
\author{Yibo~Wu\IEEEauthorrefmark{1}\IEEEauthorrefmark{2},
        Ulf~Gustavsson\IEEEauthorrefmark{1},
        Alexandre~Graell~i~Amat\IEEEauthorrefmark{2}, and
        Henk~Wymeersch\IEEEauthorrefmark{2}\\
        \IEEEauthorrefmark{1}Ericsson Research, Gothenburg, Sweden\\
        \IEEEauthorrefmark{2}Chalmers University of Technology, Gothenburg, Sweden
        }
\maketitle

\begin{abstract}
Tracking the nonlinear behavior of an RF \ac{pa} is challenging. To tackle this problem, we build a connection between residual learning and the \ac{pa} nonlinearity, and propose a novel residual neural network structure, referred to as the \ac{r2tdnn}. Instead of learning the whole behavior of the \ac{pa}, the \ac{r2tdnn} focuses on learning its nonlinear behavior by adding identity shortcut connections between the input and output layer. In particular, we apply the \ac{r2tdnn} to digital predistortion and measure experimental results on a real \ac{pa}. Compared with  neural networks recently proposed by Liu \textit{et al}. and Wang \textit{et al}., the \ac{r2tdnn} achieves the best linearization performance in terms of normalized mean square error and adjacent channel power ratio with less or similar computational complexity. Furthermore, the \ac{r2tdnn} exhibits significantly faster training speed and lower training error.

\end{abstract}

\section{Introduction}
\Ac{5g} wireless systems pose significant challenges to the performance of the \ac{rf} \acf{pa}~\cite{kelly2016preparing}. High-frequency and high-bandwidth signals suffer severe distortions from the nonlinear behavior of the \ac{pa}, which increases the need for highly linear \acp{pa}. Meanwhile, the increasing number of antennas and base-stations require a large number of \acp{pa}, which greatly increases the stress on power consumption, so the power efficiency of the \acp{pa} is also crucial. 

In practice, the linearity and efficiency of the \ac{pa} becomes a trade-off when both need to be satisfied. This trade-off has triggered intensive research over the past decades~\cite{eun1997new, kim2001digital, GMP_2006}. These works aim to preserve the \ac{pa} linearity at the high output power region by using \acf{dpd}, a well-known technique to compensate for the \ac{pa} nonlinearity. \ac{dpd} performs an inverse nonlinear operation before the \ac{pa}. This inverse operation can be represented by a parametric model, whose accuracy determines the \ac{dpd} performance. Conventionally, Volterra series based models~\cite{eun1997new}, such as \ac{mp}~\cite{kim2001digital} and \ac{gmp}~\cite{GMP_2006}, have been widely used for \ac{dpd} because of their high accuracy. In these models, the behavior of the \ac{pa} is represented by a set of Volterra kernels with different nonlinear orders where each kernel also considers memory effects, i.e., past inputs that influence the current output. These memory effects are due to the frequency-dependent behavior of the \ac{pa}~\cite{pedro2005comparative}.  However, the performance of Volterra-based models is limited for severely nonlinear \acp{pa} even if high-order kernels are used because of the high estimation error for high-order kernels~\cite{orcioni2014improving}.

In contrast to model-based \ac{dpd} approaches, deep learning techniques such as \acfp{nn} have recently been proposed for \ac{dpd}~\cite{isaksson2005wide, luongvinh2005behavioral, liu2004dynamic, mkadem2011physically, gotthans2014digital, wang2018augmented, hongyo2019deep, tarver2019design}. Among them, the \ac{mlp} is the most commonly chosen type of \acp{nn} for \ac{dpd}~\cite{liu2004dynamic, mkadem2011physically, gotthans2014digital, wang2018augmented, hongyo2019deep, tarver2019design} because of the simple implementation and training algorithm. Based on the \ac{mlp},  \cite{liu2004dynamic} proposed a \acf{rvtdnn} that separates the complex-valued signal into real in-phase and quadrature components to use a simple real-valued training algorithm. Furthermore, to consider  memory effects of the \ac{pa}, the input layer of the \ac{rvtdnn} is fed by both the current instantaneous input and the inputs at previous time instants. To improve the performance of the \ac{rvtdnn}, many variants have been studied~\cite{mkadem2011physically, gotthans2014digital, wang2018augmented}, which add more components to the input layer, such as previous samples of the output signal~\cite{mkadem2011physically}, future samples of the input signal~\cite{gotthans2014digital}, or envelope terms (e.g., amplitude) of the input signal~\cite{wang2018augmented}. However, while these additional components have been shown to improve performance, they also significantly increase the network complexity, which pushes more pressure on the power consumption of \ac{dpd}. \cite{tarver2019design} considered a different approach to connect the input and output layer by a linear bypass, which makes the \ac{nn} focus on the nonlinear relation. However, this approach is infeasible for a memory input, which limits its performance on \acp{pa} with memory. Moreover, the performance comparison between \ac{nn} with and without shortcuts for \ac{dpd} is not discussed in~\cite{tarver2019design}.

In this paper, we build a connection between  \textit{residual learning} and the \ac{pa}. We then propose a residual \ac{nn},  referred to as~\acf{r2tdnn} to learn the nonlinear behavior of the \ac{pa}. Unlike \ac{rvtdnn}~\cite{liu2004dynamic} and its variants~\cite{mkadem2011physically, gotthans2014digital, wang2018augmented} that learn the \ac{pa} linear and nonlinear behaviors  jointly, the proposed \ac{r2tdnn} learn them separately. Specifically, the \ac{pa} nonlinear behavior is learned by its inner layers, and the linear behavior is added at the end of the inner layers using \textit{identity shortcuts} between the input and output layer. The identity shortcuts introduce no new parameters as well as negligible computational complexity (one element-wise addition). Unlike~\cite{tarver2019design}, which excludes memory inputs, the \ac{r2tdnn} considers memory inputs by applying identity shortcuts between two neurons of the current instantaneous input-output, which also solve the dimension difference between the input and output layer.   We apply the proposed~\ac{r2tdnn} to \ac{dpd}. Experimental results on a real \ac{pa} show that the proposed \ac{r2tdnn} for \ac{dpd} achieves a better linearization performance as well as a faster training rate than the \ac{rvtdnn} in~\cite{liu2004dynamic} and a variant of it in~\cite{wang2018augmented} with similar computational complexity. 

\section{System Model}\label{section:model} 

\begin{figure}[tbp]
    \centering
    \includegraphics[width=0.75\linewidth]{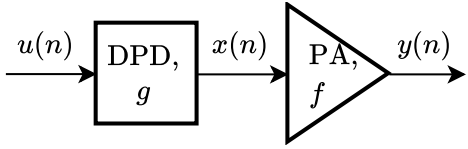}
    \caption{Behavior relation between \ac{dpd} and the \ac{pa}. The  power  gain  of  the  PA  is  normalized  for  simplicity. To compensate the \ac{pa} nonlinear behavior $f$ before the saturation point, \ac{dpd} performs an inverse operation $g$.}
    \label{fig:dpd_pa_diagram}
\end{figure}
\subsection{PA behavior and DPD}
The \ac{pa} behaves as a nonlinear system that exhibits static nonlinearity and memory effects. The latter is more obvious in a wideband scenario because of the frequency-dependent gain and phase shift between the input and output signal~\cite{939917}. Memory effects are exhibited in the time domain, which means that the \ac{pa} output at any time instant is a function of the current instantaneous input and previous inputs. To take into account memory effects, we consider the \ac{pa} as a function $f$: $\mathbb{R}^{L}\rightarrow \mathbb{R}$ with input and output signals $x(n)$ and $y(n)$ for $n\in \mathbb{Z}$, and input memory length $L$. The input-output relation of the \ac{pa} can be expressed as
\begin{equation}
    y(n) = f(x(n-L),\ldots, x(n)).
    \label{eq:pa_in_out_1}
\end{equation}
Meanwhile, the \ac{dpd} is viewed as a function $g$: $\mathbb{R}^{L_1+L_2}\rightarrow \mathbb{R}$, with delayed and advanced memory length $L_1$ and $L_2$, and input signal $u(n)$ for $n\in \mathbb{Z}$, given by
\begin{equation}
    x(n) = g(u(n-L_1),\ldots, u(n+L_2)).
    \label{eq:dpd_in_out}
\end{equation}
As shown in Fig.~\ref{fig:dpd_pa_diagram}, \ac{dpd} is placed before the \ac{pa} so as to cancel the distortion introduced by the \ac{pa}. Assuming an ideal \ac{dpd} cancellation, i.e., the \ac{dpd} perfectly compensates for the distortion introduced by the \ac{pa}, we then have the ideal input-output relation of the \ac{dpd}--\ac{pa} system by substituting \eqref{eq:dpd_in_out} into \eqref{eq:pa_in_out_1}, with $\boldsymbol{u}_{n-L} = [u(n-L-L_1),\ldots,u(n-L+L_2)]^{\mathsf{T}}$,
\begin{equation}
    y(n) = f(g(\boldsymbol{u}_{n-L}),\ldots, g(\boldsymbol{u}_{n})) = u(n).
    \label{eq:dpd-pa-ideal}
\end{equation}
In this case, the cascaded \ac{dpd}--\ac{pa} system is distortion-free.

However, an ideal \ac{dpd} cancellation is infeasible in practice because of the saturation region and other non-deterministic factors such as noise. To minimize the distortion at the output of the \ac{pa}, various behavioral models explore deterministic functions to approximate $g$ so as to make the \ac{dpd}--\ac{pa} system as linear as possible. Let $\hat{g}$ denote the approximated \ac{dpd} function, which turns the distortion-free output $u(n)$ in~\eqref{eq:dpd-pa-ideal} to a biased \ac{pa} output $\hat{u}(n)$,
\begin{align}
    y(n) &= f(\hat{g}(\boldsymbol{u}_{n-L}),\ldots, \hat{g}(\boldsymbol{u}_{n})) = \hat{u}(n).
    \label{eq:in-out-actual}
\end{align}
To reduce the output bias, a highly accurate model of $g$ is crucial.

\subsection{Generalized Memory Polynomial (GMP)}\label{subsec:GMP}
A popular form of a nonlinear, causal, and finite-memory system (e.g., the \ac{pa}) is described by the Volterra series because of good precision. To ease the high complexity of Volterra series, many simplified Volterra models have been investigated in the literature~\cite{eun1997new, kim2001digital, GMP_2006}. In particular, the \ac{gmp}~\cite{GMP_2006} behavioral model has been shown to outperform many other models in terms of accuracy versus complexity~\cite{5460970}. 

Assuming a \ac{gmp} model with memory depth $M$, nonlinear order $P$, cross-term length $G$, and input signal $x_{\text{in}}(n)$ at time $n$, the output of the \ac{gmp} at time $n$, $\hat{y}_{\text{out}}(n)$, gives an estimation of the actual output $y_{\text{out}}(n)$ as~\cite{GMP_2006}
\begin{equation}
\begin{aligned}
        \hat{y}_{\text{out}}(n) =& \sum_{p=0}^{P-1}\sum_{m=0}^{M}a_{pm}x_{\text{in}}(n-m)|x_{\text{in}}(n-m)|^{p}\\
    &+\sum_{p=1}^{P-1}\sum_{m=0}^{M}\sum_{g=1}^{G} (b_{pmg}x_{\text{in}}(n-m)|x_{\text{in}}(n-m-g)|^{p} \\
    &+ c_{pmg}x_{\text{in}}(n-m)|x_{\text{in}}(n-m+g)|^{p})
    \label{eq:GMP-formula}
\end{aligned}
\end{equation}
where $a_{pm}$, $b_{pmg}$, and $c_{pmg}$ are complex-valued coefficients. Assuming a total number of coefficients $J$ and total number of input samples $N$, all coefficients can be collected into a $J\times1$ vector $\boldsymbol{w}$. Each element of $\boldsymbol{w}$ corresponds to a $N\times 1$ signal, e.g., coefficient $a_{32}$ corresponds to the $N$ samples signal $x_{\text{in}}(n-2)|x_{\text{in}}(n-2)|^{3}$. Therefore, we can collect these $N\times 1$ input signals into the $N\times J$ matrix $\boldsymbol{X}_{\text{in}}$.
Then, \eqref{eq:GMP-formula} can be rewritten in matrix form as
\begin{equation}
    \hat{\boldsymbol{y}}_{\text{out}} = \boldsymbol{X}_{\text{in}}\boldsymbol{w}.
    \label{eq:GMP-formula-matrix}
\end{equation}

To solve for $\boldsymbol{w}$, the least squares algorithm is commonly used by minimizing the \ac{mse} between the estimation $\hat{\boldsymbol{y}}_{\text{out}}$ and the observation $\boldsymbol{y}_{\text{out}}$, which gives a solution for $\boldsymbol{w}$,
\begin{equation}
    \boldsymbol{w} = (\boldsymbol{X}_{\text{in}}^\mathsf{H}\boldsymbol{X}_{\text{in}})^{-1}\boldsymbol{X}_{\text{in}}^\mathsf{H}\boldsymbol{y}_{\text{out}},
\end{equation}
where $\mathsf{H}$ denotes Hermitian.

In a real-time scenario, the running complexity of \ac{dpd} substantially restricts the system. Assuming $G<M+1$, reference~\cite{5460970} computes the running complexity of \ac{gmp}, $C_{\text{GMP}}$, for each input sample in terms of the number of \ac{flops},
\begin{equation}
\begin{aligned}
    C_{\text{GMP}} =& 8\left((M+1)(P+2PG) - \frac{G(G+1)}{2}(P-1)\right)\\
    &+ 10 + 2P + 2(P-1)G + 2P\text{min}(G,M).
    \end{aligned}
\end{equation}

\subsection{Inverse Structure to identify \ac{dpd} coefficients}
Before a behavioral model (e.g., the \ac{gmp} model) is used to represent the \ac{dpd} function $g$, we need to identify its coefficients. Since the \ac{dpd} optimal output signal $x(n)$ is unknown, we cannot directly identify coefficients of a model using $u(n)$ and $x(n)$. Alternatively, we can use an inverse structure, the \ac{ila}~\cite{eun1997new}, to indirectly identify \ac{dpd} parameters.  First, an \textit{inverse} \ac{pa} model (also known as \textit{post-distorter}) is identified using the \ac{pa} output signal $y(n)$ as the input and the \ac{pa} input signal $u(n)$ as the output. Once the the post-distorter is identified, its coefficients are copied to to an identical model (known as \textit{pre-distorter}) which is then used as the \ac{dpd} function $g$. 

Although the learned post-distorter is not an optimal solution, \ac{ila} is still the most used identification method because of simple implementation and good performance. In this paper, we consider the \ac{ila} to identify the parameters of a \ac{dpd} model.

\section{Proposed Residual Real-Valued Time-Delay Neural Network} \label{section:R2TDNN}
In this section we build a connection between the residual learning and the \ac{pa} behavior, and then propose a residual \ac{nn} to learn the nonlinear behavior of the \ac{pa}.
\subsection{Residual learning on the PA}
The \ac{pa} behavior consists of a linear and a nonlinear component. If we extract the linear relation, the input-output relation of the \ac{pa}~\eqref{eq:pa_in_out_1} can be rewritten as
\begin{equation}
\begin{aligned}
    y(n) = x(n) + \underbrace{f(x(n-L),\ldots,x(n)) - x(n)}_{=h(x(n-L),\ldots,x(n))}.
    \label{eq:residual_f}
\end{aligned}
\end{equation}
Here, let us refer to $f(x(n-L),\ldots,x(n))$ as the original function to be learned by the \ac{nn}, and the last two terms on the right-hand side of~\eqref{eq:residual_f}, i.e., $f(x(n-L),\ldots,x(n)) - x(n)$, as the residual function, which is denoted by $h(x(n-L),\ldots,x(n))$. 

In the field of image recognition, learning a residual function has been shown to be more effective than learning its corresponding original function~\cite{ResNet_Kaiming}. Therefore, we hypothesize that learning the nonlinear behavior of the \ac{pa} is easier than learning the whole behavior. We then propose a residual learning \ac{nn} to learn the \ac{pa} behavior, referred to as~\ac{r2tdnn}. Unlike the \ac{rvtdnn}~\cite{liu2004dynamic} and its variants~\cite{mkadem2011physically, gotthans2014digital, wang2018augmented} that learn the whole input-output relation of the \ac{pa} jointly, i.e., learn the original function $f(x(n-L),\ldots,x(n))$, the proposed \ac{r2tdnn} learns it separately as in \eqref{eq:residual_f}. In particular, the residual function $h(x(n-L),\ldots,x(n))$, i.e., the \ac{pa} nonlinear behavior, is learned by inner layers, and $x(n)$, i.e., the \ac{pa} linear behavior, is then added to the output of the inner layers by using \textit{shortcut} connections between input and output layers. Specifically, we adopt the identity shortcut, which performs an \textit{identity} mapping between connected layers and introduces no extra parameters.  The details of the identity shortcut in the \ac{r2tdnn} are described in the next subsection.

\subsection{Architecture}
\begin{figure}[t]
    \centering
    \includegraphics[width=1\linewidth]{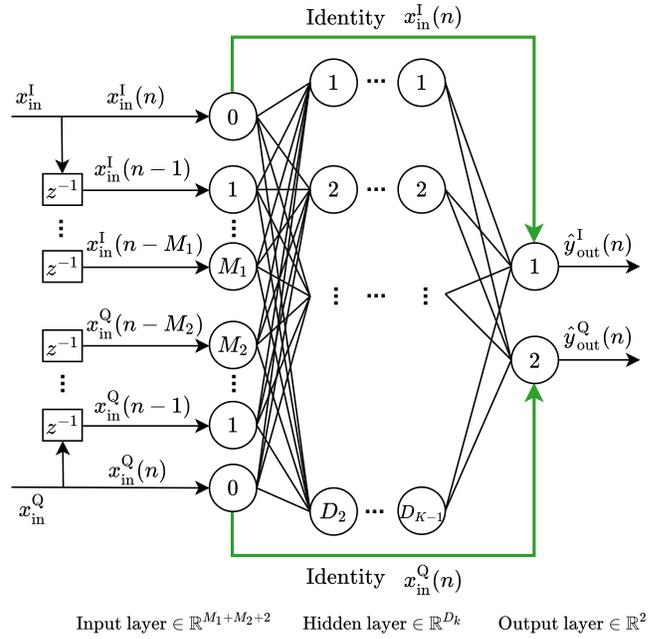}
    \caption{Architecture of the proposed \ac{r2tdnn}. Fed by the real in-phase and quadrature components of the input signal, $x_{\text{in}}^{\text{I}}$ and $x_{\text{in}}^{\text{Q}}$, the \ac{r2tdnn} gives the I and Q output signal estimations $\hat{y}_{\text{out}}^{\text{I}}$ and $\hat{y}_{\text{out}}^{\text{Q}}$. }
    \label{fig:R2TDNN_diagram}
\end{figure}
The architecture of the \ac{r2tdnn} is shown in Fig. \ref{fig:R2TDNN_diagram}. Based on the \ac{mlp}, the \ac{r2tdnn} consists of $K$ layers. The number of neurons of layer $k$ is denoted by $D_k$. The input vector of layer $k$ is denoted by $\boldsymbol{z}_k \in \mathbb{R}^{D_{k-1}}$, which is also the output of layer $k-1$. We denote the weight matrix and bias vector of layer $k$ by $\boldsymbol{W}_{k} \in \mathbb{R}^{D_{k} \times D_{k-1}}$ and $\boldsymbol{b}_{k} \in \mathbb{R}^{D_{k}}$, respectively.
We consider a real-valued \ac{mlp}, so the complex-valued input signal $x_{\text{in}}(n) = x_{\text{in}}^{\text{I}}(n) + j x_{\text{in}}^{\text{Q}}(n)$ at time instant $n$ is separated into real in-phase and quadrature components, $x_{\text{in}}^{\text{I}}(n)$ and $x_{\text{in}}^{\text{Q}}(n)$, respectively. To learn memory effects of the \ac{pa}, the input signal of the first layer is formed by tapped delay lines, where each delay operator $z^{-1}$ yields one time instant delay, e.g., $x_{\text{in}}^{\text{I}}(n)$ to $x_{\text{in}}^{\text{I}}(n-1)$. We consider memory length $M_1$ and $M_2$ for the I and Q input components, respectively. Thus, the input signal of the first layer at time instant $n$ is given by
\begin{equation}
\begin{aligned}
    \boldsymbol{z}_{\text{1}}(n) = [&x_{\text{in}}^{\text{I}}(n), x_{\text{in}}^{\text{I}}(n-1),...,x_{\text{in}}^{\text{I}}(n-M_1), \\ &x_{\text{in}}^{\text{Q}}(n), x_{\text{in}}^{\text{Q}}(n-1), ..., x_{\text{in}}^{\text{Q}}(n-M_2)],
    \label{eq:input_vector_R2TDNN}
    \end{aligned}
\end{equation}
which yields $(M_1+M_2+2)$ number of neurons for the first layer.  When $M_1=M_2=0$, the network neglects memory. 

The layer $k-1$ and $k$ are fully connected as
\begin{equation}
    \boldsymbol{z}_{k+1} = \sigma(\boldsymbol{W}_{k} \boldsymbol{z}_{k} +\boldsymbol{b}_{k}),
\end{equation}
where $\sigma$ is the activation function. 
The output of the last layer is a $2\times1$ vector which corresponds to the in-phase and quadrature output signal estimations $\hat{y}_{\text{out}}^{\text{I}}(n)$ and $\hat{y}_{\text{out}}^{\text{Q}}(n)$ of the actual complex-valued output signal $y_{\text{out}}(n)$. To output a full range of values, the output layer is considered as a linear layer with no activation function. More importantly, we add the identity shortcut between the input and output layers. Unlike other shortcuts that fully connect two layers, as in~\cite{ResNet_Kaiming}, here the identity shortcut connection is performed between neurons. Only the two neurons fed by the current time instant input signal, i.e., $x_{\text{in}}^{\text{I}}(n)$ and $x_{\text{in}}^{\text{Q}}(n)$, are
connected to the output neurons. Therefore, the output of layer $K$ can be written as
\begin{equation}
    [\hat{y}_{\text{out}}^{\text{I}}(n), \hat{y}_{\text{out}}^{\text{Q}}(n)] = [x_{\text{in}}^{\text{I}}(n), x_{\text{in}}^{\text{Q}}(n)] + \boldsymbol{W}_K \boldsymbol{z}_{K} + \boldsymbol{b}_K.
    \label{eq:r2tdnn_output_1}
\end{equation}
Note that the last two terms on the right hand side of~\eqref{eq:r2tdnn_output_1} represents the residual function $h$ in~\eqref{eq:residual_f}, whereas the identity shortcut accounts for the linear part.

\subsection{Computation Complexity}
The identity shortcut connection introduces no new parameters to the \ac{nn}, and only two element-wise additions are added to the running complexity. All multiplications and additions are performed between real values, which accounts for one FLOP according to~\cite[Table I]{5460970}.

The number of \ac{flops} needed for the \ac{r2tdnn} is
\begin{equation}
    C_{\text{R2TDNN}} = 2\sum_{k=1}^{K-1}D_k D_{k+1} + 2,
    \label{eq:C_R2TDNN_1}
\end{equation}
where the first term is the number of \ac{flops} for multiplication and addition operations, and the $2$ is for two addition operations contributed by the two identity shortcuts. 

\subsection{\ac{r2tdnn} on \ac{dpd}}
The parameters of the \ac{r2tdnn} can be learned through the back-propagation algorithm by minimizing the \ac{mse} between the prediction $\hat{y}_{\text{out}}(n)$ and observation $y_{\text{out}}(n)$,
\begin{equation}
    (\boldsymbol{W}^*,\boldsymbol{b}^*) = \text{arg }\underset{\boldsymbol{W},\boldsymbol{b}}{\text{min }}\mathbb{E}[(y_{\text{out}}(n)-\hat{y}_{\text{out}}(n))^2],
\end{equation} 
where $\mathbb{E}[\cdot]$ denotes the expectation.
Specifically, when the \ac{r2tdnn} is used as \ac{dpd}, its parameters can be identified using the \ac{ila}, where the \ac{pa} output $y(n)$ and input $x(n)$ are fed to the \ac{r2tdnn} as input $x_{\text{in}}$ and output $y_{\text{out}}$, respectively.

\section{Experimental results}\label{section:results}
We give experimental results of applying different behavioral models to \ac{dpd} on a real \ac{pa}.

\subsection{Evaluation Metrics and Measurement Setup}\label{section:metric_Weblab}
\subsubsection{Evaluation Metrics}
To evaluate the performance of \ac{dpd}, the distortion level of the \ac{pa} output signal is generally measured by the \ac{nmse} between the \ac{pa} output signal $y(n)$ (with gain normalization) and \ac{dpd} input signal $u(n)$, and the \ac{acpr} of $y(n)$.

The \ac{nmse} is defined as
\begin{equation}
    \text{NMSE} = \frac{ \sum\limits_{n}|y(n) - u(n)|^2}{\sum\limits_{n}|u(n)|^2}.
    \label{eq:NMSE}
\end{equation}
Although the \ac{nmse} measures the all-band distortion, it can be used to represent the in-band distortion as the power of out-of-band distortion is negligible compared to the in-band distortion.

The \ac{acpr} measures the ratio of the out-of-band leakage to the in-band power, and is defined as
\begin{equation}
    \text{ACPR} = \frac{\int_{\text{adj.}} |Y(f)|^2\text{d}f}{\int_{\text{ch.}} |Y(f)|^2\text{d}f},
    \label{eq:ACPR}
\end{equation}
where $Y(f)$ denotes the Fourier transform of the \ac{pa} output signal. The integration in the numerator and denominator are done over the adjacent channel (the lower or upper one with a larger leakage) and the main channel, respectively.

\subsubsection{Measurement Setup}\label{section:RF_Weblab}
\begin{figure}[tbp]
    \centering
    \includegraphics[width=0.9\linewidth]{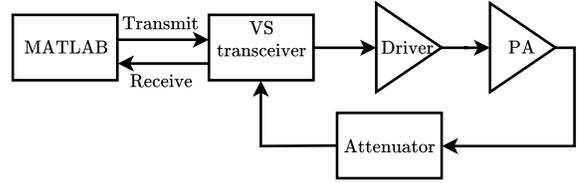}
    \caption{Block diagram of the measurement setup. The RF WebLab is remotely accessed by the MATLAB which transmits and receives the pre-distorted and measured signals, respectively. }
    \label{fig:RF_weblab_diagram}
\end{figure}
The experimental setup is based on the RF WebLab\footnote{RF WebLab is a PA measurement setup that can be remotely accessed at \url{www.dpdcompetition.com}}~\cite{landin2015weblab}. Fig. \ref{fig:RF_weblab_diagram} illustrates how it interacts with the hardware and \ac{dsp} algorithms, e.g., \ac{dpd}. In RF WebLab, a \ac{vst} (PXIe-5646R VST) transmitter generates analog signals based on the digital signal from MATLAB. Signals are then sent to the Gallium Nitride \ac{pa} DUT (Cree CGH40006-TB) driven by a 40 dB linear driver. Then, after a 30 dB attenuator,  the \ac{vst} receiver obtains the \ac{pa} output signals, and eventually measurements are sent back to the MATLAB. 

We then apply the proposed \ac{r2tdnn}, \ac{gmp}~\cite{GMP_2006}, \ac{rvtdnn}~\cite{liu2004dynamic}, and \ac{arvtdnn}~\cite{wang2018augmented} to \ac{dpd} with the RF WebLab setup. The learning architecture for all \ac{dpd} models is the \ac{ila}~\cite{eun1997new} because of simple implementation. To identify \ac{dpd} coefficients, \ac{gmp} adopts the least squares algorithm, while \ac{rvtdnn}, \ac{arvtdnn}, and \ac{r2tdnn} use the back-propagation algorithm with the \ac{mse} loss function. We choose Adam~\cite{kingma2014adam} as the optimizer with a mini-batch size of $256$ and a learning rate of $0.001$. The activation function is the leaky \ac{relu} with a slope of $0.01$ for a negative input.

The input signal $u(n)$ is an \ac{ofdm} signal with length $10^6$, sampling rate $200$ MHz, and signal bandwidth $10$ MHz. We consider a $50$ $\Omega$ \ac{pa} load impedance. The measured saturation point and measurement noise variance of the \ac{pa} are $24.1$ V ($\approx 37.6$ dBm) and $0.0033$, respectively. To test the \ac{dpd} performance on the \ac{pa} nonlinear region, we consider an average output power of the \ac{pa} output signal of $25.36$ dBm, where the corresponding theoretical minimum \ac{nmse} is $-40.17$ dB according to~\cite[Eq. (10)]{chani2018lower}, and the simulated minimum \ac{acpr} is $-50.1$ dBc.\footnote{The simulated minimum \ac{acpr} represents the \ac{acpr} of the ideal linear \ac{pa} output signal.}

\subsection{Results}\label{section:results_sub}
\subsubsection{Performance versus Complexity}\label{subsec:complexity}
\begin{figure}[t]
    \centering
    \includegraphics[width=0.95\linewidth]{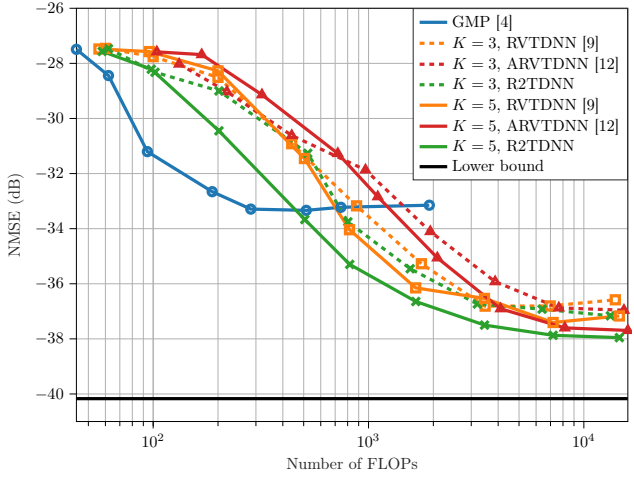}
    \caption{NMSE as a function of the number of \ac{flops}. $M_1=M_2=3$. The markers for \ac{rvtdnn}~\cite{liu2004dynamic}, \cite{wang2018augmented}, and \ac{r2tdnn} correspond to different $D_k$. The markers for \ac{gmp} correspond to different values of $P$, $M$, and $G$. The lower bound represents the theoretical minimum \ac{nmse} that can be achieved for this \ac{pa} at an average output power $25.36$ dBm.}
    \label{fig:Flop_nmse}
\end{figure}

\begin{figure}[t]
    \centering
    \includegraphics[width=0.95\linewidth]{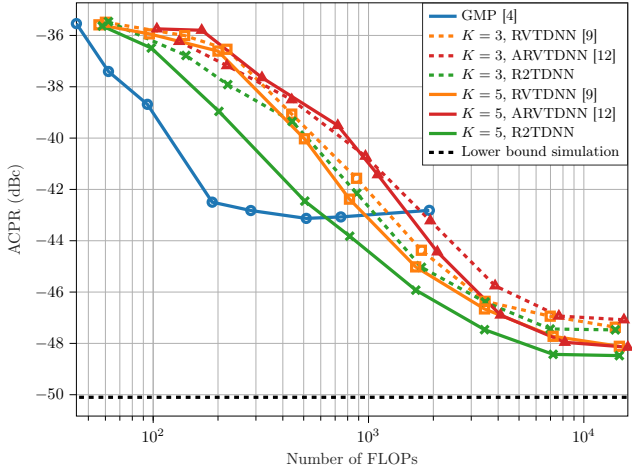}
    \caption{ACPR as a function of the number of \ac{flops}. $M_1=M_2=3$. The markers for \ac{rvtdnn}~\cite{liu2004dynamic}, \ac{arvtdnn}~\cite{wang2018augmented}, and \ac{r2tdnn} correspond to different $D_k$. The markers for \ac{gmp} correspond to different values of $P$, $M$, and $G$. The lower bound simulation represents the simulated minimum \ac{acpr} that can be achieved for this \ac{pa} at an average output power $25.36$ dBm.}
    \label{fig:Flop_acpr}
\end{figure}

\begin{table}[t]
\centering
\vspace{0.5\baselineskip}
\caption{\ac{nmse} and \ac{acpr} results of the \ac{rvtdnn}~\cite{liu2004dynamic}, \ac{arvtdnn}~\cite{wang2018augmented}, and \ac{r2tdnn} for the convergence in Fig.~\ref{fig:train_val}. $K=5$, $D_2=D_3=D_4=9$.}
\begin{center}
\begin{tabularx}{\linewidth}{X*{4}c}
\hline
       & Num. FLOPs & NMSE {[}dB{]} & ACPR {[}dBc{]} \\ \hline
RVTDNN~\cite{liu2004dynamic} &  $504$ & $-31.5$        & $-40.0$         \\
ARVTDNN~\cite{wang2018augmented} &  $720$ & $-31.3$        & $-39.5$ \\
R2TDNN & $506$ &  $\mathbf{-33.7}$      & $\mathbf{-42.5}$  
\end{tabularx}
\end{center}
\label{tab:training_comp_setup}
\end{table}

\begin{figure}[!h]
    \subcaptionbox{Training}
    {
    \includegraphics[width=0.95\linewidth]{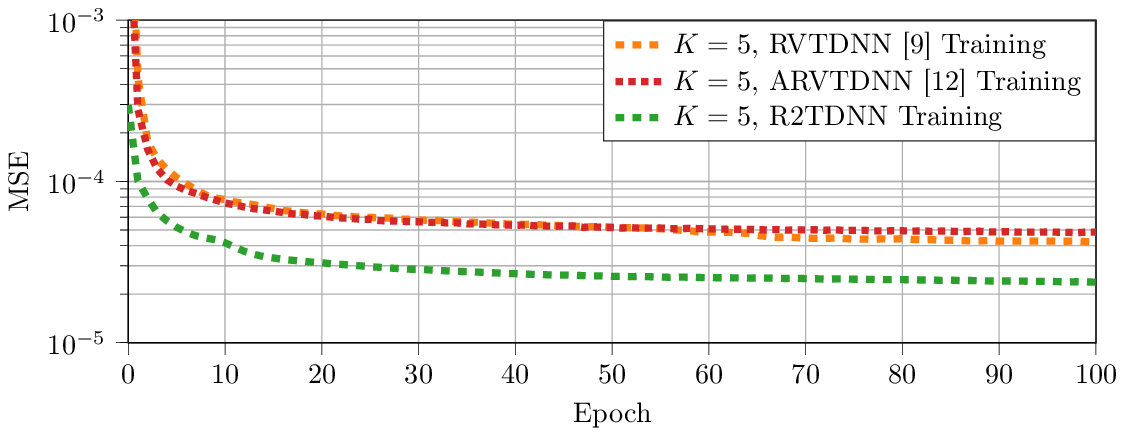}
    \vspace*{-0.5\baselineskip}
    \label{fig:_train}
    }\hfill
    \subcaptionbox{Validation}
        {
        \includegraphics[width=0.96\linewidth]{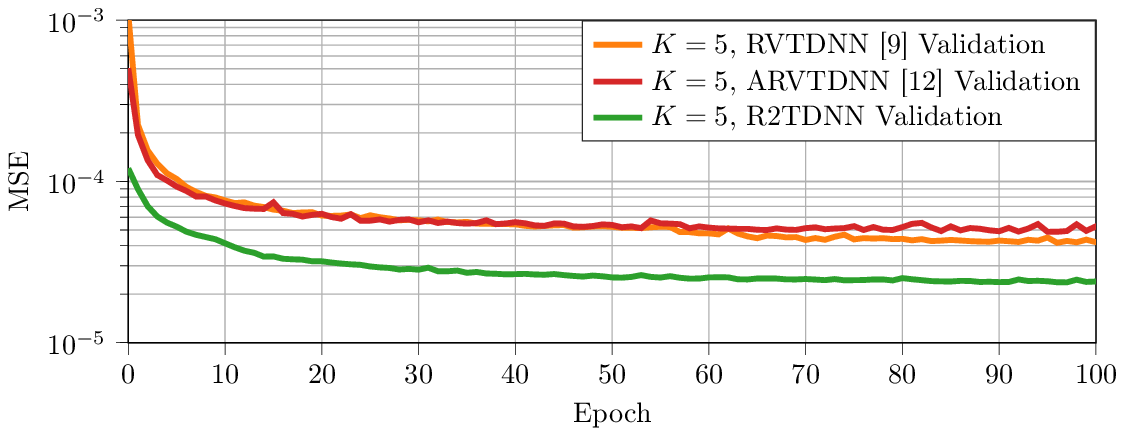}
        \vspace*{-0.5\baselineskip}
        \label{fig:_val}
        }
    \caption{Training and validation errors during the training process of the \ac{rvtdnn}~\cite{liu2004dynamic}, \ac{arvtdnn}~\cite{wang2018augmented}, and proposed~\ac{r2tdnn}. $K=5$ and $D_2=D_3=D_4=9$.}
    \label{fig:train_val}
\end{figure}

Fig. \ref{fig:Flop_nmse} and Fig. \ref{fig:Flop_acpr} show the \ac{nmse} results versus the total number of \ac{flops} for the \ac{gmp}~\cite{GMP_2006}, \ac{rvtdnn}~\cite{liu2004dynamic}, \ac{arvtdnn}~\cite{wang2018augmented}, and the proposed \ac{r2tdnn}. For the \ac{rvtdnn} and \ac{r2tdnn}, we consider two scenarios with one and three hidden layers, i.e., $K\in\{3,5\}$. We also plot the results of the \ac{arvtdnn}~\cite{wang2018augmented} for $K=5$, where we consider three augmented envelop terms of the signal (amplitude and its square and cube)~\cite[Tab. II]{wang2018augmented} at the input layer. A proper number of memory length is related to the \ac{pa} characteristics and input signal bandwidth, and here we choose identical input memory $M_1=M_2=3$ for \ac{rvtdnn}, \ac{arvtdnn}, and \ac{r2tdnn}. Meanwhile, they use the same number of neurons for each hidden layer. The number of \ac{flops} increases as the number of neurons for each hidden layer increases. For the \ac{gmp} (blue circle markers), we select the best results  with respect to the number of \ac{flops} based on an exhaustive search of different values of $P$, $M$, and $G$.

Although the \ac{gmp} model achieves better \ac{nmse} for a number of \ac{flops} $< 500$, the performance flattens around $-33.41$ dB. The proposed \ac{r2tdnn} allows to reach lower \ac{nmse} (down to $-38.0$ dB) for a number of \ac{flops} $> 500$, i.e., the \ac{r2tdnn} yields more accurate compensation\textemdash it can find a better inverse behavior of the \ac{pa}. Note that the \ac{nmse} gap between the \ac{r2tdnn} and the lower bound may be due to the limitation of the \ac{ila} and some stochastic noise, e.g., phase noise. The \ac{acpr} results in Fig. \ref{fig:Flop_acpr} illustrate similar advantages of the \ac{r2tdnn} over the \ac{gmp} for a number of \ac{flops} $>600$.

For comparison, we also plot the performance of the \ac{rvtdnn} in~\cite{liu2004dynamic} and \ac{arvtdnn} in~\cite{wang2018augmented} for $K\in\{3,5\}$. We note that the \ac{arvtdnn} requires a number of \ac{flops} $>3000$ to improve the performance of \ac{rvtdnn}. However, the proposed \ac{r2tdnn} achieves lower \ac{nmse} and \ac{acpr} with respect to the \ac{rvtdnn} and \ac{arvtdnn} for a similar number of \ac{flops}.  The gain is more considerable for $K=5$ and a number of neurons per hidden layer between $5$ and $12$.

\subsubsection{Convergence speed comparison}
To further compare the \ac{rvtdnn}~\cite{liu2004dynamic}, \ac{arvtdnn}~\cite{wang2018augmented}, and \ac{r2tdnn}, we plot the training and validation errors during the training procedure in Fig.~\ref{fig:train_val}. Based on the same parameter setup in Section~\ref{subsec:complexity}, we select $K=5$ and $D_2=D_3=D_4=9$. The corresponding number of \ac{flops}, \ac{nmse}, and \ac{acpr} are given in Table \ref{tab:training_comp_setup}. Compared to the \ac{rvtdnn} and \ac{arvtdnn}, the \ac{r2tdnn} exhibits significantly faster training convergence rate, and eventually achieves lower training and validation errors. This verifies the effectiveness of the proposed residual learning on \ac{dpd}.

\section{Conclusion}\label{section:conclusion}
We applied residual learning to facilitate the learning problem of the \ac{pa} behavior, and proposed a novel \ac{nn}-based \ac{pa} behavioral model, named \ac{r2tdnn}. By adding shortcuts between the input and output layer, the proposed \ac{r2tdnn} focus on learning the \ac{pa} nonlinear behavior instead of learning its whole behavior. We applied different behavioral models to \ac{dpd} and evaluated the performance on a real \ac{pa}. Results show that the proposed \ac{r2tdnn} achieves lower \ac{nmse} and \ac{acpr} than the \ac{rvtdnn} and \ac{arvtdnn} previously proposed in the literature with less or similar computational complexity. Furthermore, it has a faster training convergence rate during the training procedure.
 \balance

\bibliographystyle{./bibliography/IEEEtran}
\bibliography{main}
\end{document}